\documentclass[10pt, conf]{IEEEtran}
\ifCLASSINFOpdf
   \usepackage[pdftex]{graphicx}
   \graphicspath{{../figures/}{../plots/}}
\else
   \usepackage[dvips]{graphicx}
   \graphicspath{{../figures/}{../plots/}}
\fi
\usepackage[cmex10]{amsmath}
\usepackage{amssymb,verbatim,bm}
\usepackage{amsthm}
\usepackage[tight,footnotesize]{subfigure}
\newcommand{\B}{\mathcal{B}}
\newcommand{\C}{\mathcal{C}}
\newcommand{\F}{\mathcal{F}}
\newcommand{\myprob}[1]{\mathsf{P}\left\{#1\right\}}
\newcommand{\myexp}[1]{\mathsf{E}\left[#1\right]}
\newtheorem{lemma}{Lemma}

\newcommand{\nn}{\nonumber\\}

\usepackage{color}
\begin{document}
\title{Utility Optimal Coding for Packet Transmission over Wireless
       Networks -- Part II:\\ Networks of Packet Erasure Channels}
		
		\author{K.~Premkumar, Xiaomin Chen, and Douglas J.~Leith\\
        Hamilton Institute, National University of Ireland, Maynooth,
		Ireland\\
		E--mail: \{Premkumar.Karumbu, Xiaomin.Chen, Doug.Leith\}@nuim.ie

        \thanks{This work is supported by Science Foundation Ireland
		under Grant No. 07/IN.1/I901.}
       }
\maketitle
\thispagestyle{empty}
\pagestyle{empty}

\begin{abstract}
We define a class of multi--hop erasure networks that approximates a
wireless multi--hop network. The network carries unicast flows for
multiple users, and each information packet within a flow is required to
be decoded at the flow destination within a specified delay deadline.
The allocation of coding rates amongst flows/users is constrained by
network capacity. We propose a proportional fair transmission scheme
that maximises the sum utility of flow throughputs. This is achieved by
{\em jointly optimising the packet coding rates and the allocation of
bits of coded packets across transmission slots.} 
\end{abstract}


%
\IEEEpeerreviewmaketitle

\section{Introduction}
In a communication network, the network capacity is shared by a set of
flows. There is a contention for resources among the flows, which leads
to many interesting problems. One such problem, is {\em how to allocate
the resources optimally across the (competing) flows, when the physical
layer is erroneous}. Specifically, schedule/transmit time for a flow is
a resource that has to be optimally allocated among the competing flows.
In this work, we pose a network utility maximisation problem subject to 
scheduling constraints that solve a resource allocation problem. In
another work, we studied the problem of optimal resource allocation in
networks \cite{bsc}. 

\begin{figure}[t]
\centering
\includegraphics[width=75mm, height=75mm]{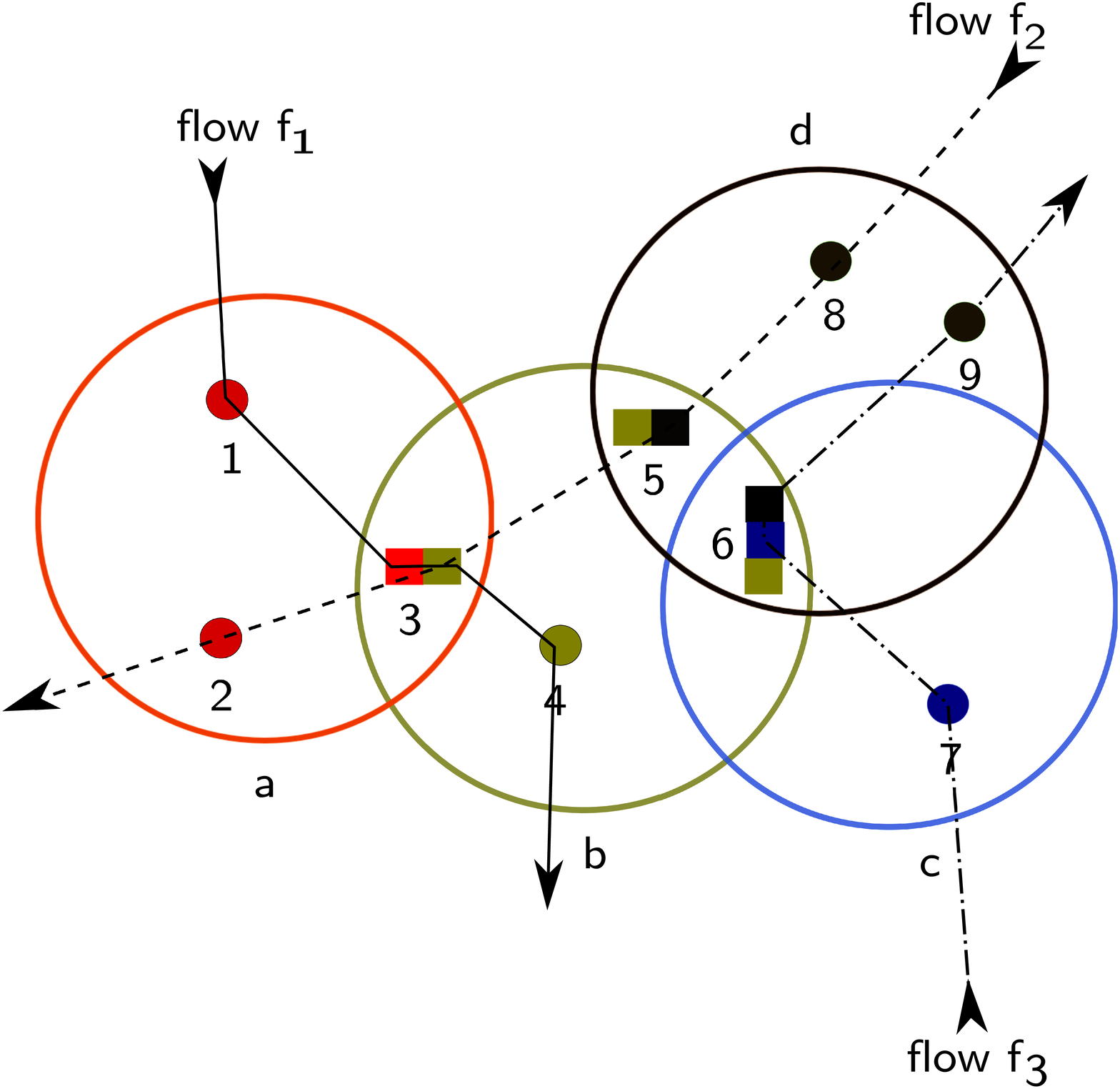}
\caption{{\bf An illustration of a wireless mesh network with 4 cells.}
Cells $a$, $b$, $c$, and $d$ use orthogonal channels CH$_1$, CH$_2$,
CH$_3$, and CH$_4$ respectively. Nodes 3, 5, and 6 are {\em bridge
nodes}. The bridge node 3 (resp. 5 and 6) is provided a time slice of
each of the channels CH$_1$ \& CH$_2$ (resp. CH$_2$ \& CH$_4$ for node 5
and CH$_2$\& CH$_3$\& CH$_4$ for node 6). Three flows $f_1, f_2$, and
$f_3$ are considered. In this example, $\C_{f_1} = \{a,b\}$, $\C_{f_2} =
\{d,b,a\}$, and $\C_{f_3} = \{c,d\}$.} 
\label{fig:mesh_network}
\end{figure}

We define a class of multi--hop erasure networks, and consider packet
communication over this class. The
network consists of a set of $C \geq 1$ cells $\C = \{1,2,\cdots,C\}$
which define the ``interference domains'' in the network. We allow
intra--cell interference (\emph{i.e} transmissions by nodes within the
same cell interfere) but assume that there is no inter--cell
interference. This captures, for example, common network architectures
where nodes within a given cell use the same radio channel while
neighbouring cells using orthogonal radio channels. Within each cell,
any two nodes are within the decoding range of each other, and hence,
can communicate with each other. The cells are interconnected using
multi--radio bridging nodes to create a multi--hop wireless network. A
multi--radio bridging node $i$ connecting the set of cells
$\B(i)=\{c_1,..,c_n\}\subset \C$ can be thought of as a set of $n$
single radio nodes, one in each cell, interconnected by a high--speed,
loss--free wired backplane (see Figure~\ref{fig:mesh_network}).

\begin{figure}
\centering
\includegraphics[width=3.8in, height=16mm]{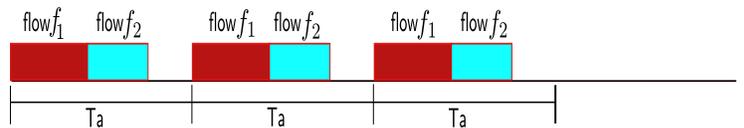}
\caption{{\bf An illustration of transmission scheme in cell $a$ of  
the network in Figure~\ref{fig:mesh_network}:} Every transmission schedule of $T_a$
time units is time--shared by nodes 1 and 3.
Note that $\phi_\Delta(f) N_f R_f$ symbols of the encoded
packet $p$ are transmitted in transmission schedule $p+\Delta$, where
$\Delta \in \{0,1,2,\cdots,n_f-1\}$. The scheduling or capacity
constraint of cell $a$ may not be tight (indicated by empty time slice
in the figure), as the rates of flows $f_1$ and
$f_2$ are governed by the whole network.}
\label{fig:transmission_scheme}
\end{figure}

Data is transmitted across this multi--hop network as a set $\F$ $=$
$\{1,2,\cdots,F\}$, $F\geq 1$ of unicast flows. The route of each flow
$f$ $\in \F$ is given by $\C_f$ $=$ $\{c_1(f), c_2(f), \cdots,
c_{\ell_f}(f)\}$, where the source node $s(f) \in c_1(f)$ and the
destination node $d(f) \in c_{\ell_f}(f)$. We assume loop--free flows
(i.e., no two cells in $\C_f$ are same).  
Figures~\ref{fig:mesh_network} and \ref{fig:transmission_scheme}
illustrate this network setup. A scheduler assigns a time slice of
duration $T_{f,c} > 0$ time units to each flow $f$ that flows through
cell $c$, subject to the constraint that $\sum_{f:c\in \C_f}
T_{f,c}\le T_c$ where $T_c$ is the period of the schedule in cell $c$.
We consider a periodic scheduling strategy (see 
Figure~\ref{fig:transmission_scheme}) in which, in each cell $c$,
service is given to the flows in a round robin fashion, and that each
flow $f$ in cell $c$ gets a time slice of $T_{f,c}$ units in every schedule.

The scheduled transmit times for flow $f$ in source cell $c_1(f)$ define
time slots for flow $f$. We assume that a new information packet arrives
in each time slot, which allows us to simplify the analysis by ignoring
queueing. Information packets of each flow $f$ at the source node $S(f)$
consist of a block of $k_f$ symbols. Each packet of flow $f$ is encoded
into codewords of length $n_f = k_f/r_f$ symbols, with coding rate
$0< {r_f}\le1$. The code employed for encoding is discussed in Section 
\ref{sec:problem_formulation}. We require sufficient transmit times 
at each cell along route $\C_f$ to allow $n_f$ coded symbols to be 
transmitted in every schedule period. Hence there is no queueing at the 
cells along the route of a flow.  
It is not apparent at this point whether it is
optimal for flow $f$ to transmit a single code--word of $n_f$ symbols or
transmit a block of $n_f$ symbols where each block carries some portions
of each of a set of coded packets.

{\bf Channel Model:}
The channel in cell $c$ for flow $f$ is considered to be a packet
erasure channel with the probability of packet erasure being
$\beta_{f,c} \in [0,1]$. Thus, the end--to--end channel for flow $f$ is
a packet erasure channel with the probability of packet erasure being
\begin{eqnarray*}
\beta_f = 1- \prod_{c \in \C_f} \left[1-\beta_{f,c}\right]
\end{eqnarray*}
Let the Bernoulli random variable $E_{f}[i]$ indicate the end--to--end
erasure seen by the $i$th block of flow $f$ (independent of the erasure
seen by other blocks) of flow $f$. Note that $E_f[i]=1$ means that the
$i$th block is erased, and $E_f[i]=0$ means that the $i$th block is
received successfully. Note that ${\sf P}\{E_f[i]=1\}= \beta_f = 1 - 
{\sf P}\{E_f[i]=0\}$.

Each packet has a deadline of $D_f$ slots, by which time it must be
decoded. Such a delay constraint is natural in applications such as
video streaming. A packet is in error if the destination fails to
decode the packet by the deadline. Letting $e_f(r_f)$ denote the error
probability that a packet fails to be decoded before its deadline, 
the expected number of information symbols successfully received is
$S_f(r_f)=k_f(1-e_f(r_f))$. Other things being equal, we expect that
decreasing $r_f$ (i.e., increasing the number of coded symbols $n_f =
k_f/r_f$ sent) decreases error probability $e_f$ and so increases $S_f$.
However, since the network capacity is limited, and is shared by
multiple flows, increasing the coded packet size $n_{f_1}$ of flow $f_1$
generally requires decreasing the packet size $n_{f_2}$ for some other
flow $f_2$. That is, increasing $S_{f_1}$ comes at the cost of
decreasing $S_{f_2}$. We are interested in understanding this
trade--off, and in analysing the optimal fair allocation of coding rates
amongst users/flows.   
 
Our main contribution is the analysis of fairness in the allocation of
coding rates between users/flows competing for limited network capacity.
In particular, we adopt a utility--fair framework, and propose a scheme
for obtaining the proportional fair allocation of coding rates,
\emph{i.e.} the allocation of coding rates that maximises
$\sum_{f\in\F}\log S_f(r_f)$ subject to network capacity constraints.
This problem, which we show in Section~III, requires solving a
non--convex optimisation problem.  Specifically,
at the physical layer, the (channel) coding rate of a flow can be
lowered (to alleviate its channel errors) only at the expense of
increasing the coding rates of other flows. Also, at the network layer,
the length of schedules of each flow should be chosen in such a way that
it maximises the network utility. Interestingly, we show in our problem
formulation that the coding rate and the scheduling are tightly coupled.
Also, we show that for a $\log$ (network) utility function (which typically gives 
proportional fair allocation of resources) the optimum rate allocation
(in general) gives unequal air--times which is quite different from the previously
known result of proportional fair allocation being the same as that of
equal air--time allocation (\cite{eq_air_time}).
This problem, which we show in Section~\ref{sec:num}, requires solving a
non--convex optimisation problem. Our work differs from the previous
work on network utility maximisation (see \cite{net-opt} and the
references therein) in the following manner. To the best of our
knowledge, this is the first work that computes the optimal coding rate
for a given scheduling (or capacity) constraints in the utility--optimal
framework.

The rest of the paper is organised as follows. In
Section~\ref{sec:problem_formulation}, we obtain a measure for the
end--to--end packet erasure, and describe the
throughput of the network. We then formulate a network utility
maximisation problem subject to constraints on the transmission schedule
lengths. In Section~III, we obtain the optimum transmission strategy and the
optimum packet--level coding rates for each flow in the network. In
Section~\ref{sec:examples}, we provide some simple examples to
illustrate our results. Due to lack of space, the proofs of various
Lemmas are omitted. 

\section{Problem Formulation}
\label{sec:problem_formulation}

The encoding has two stages.  The first stage is the encoding of each
information packet using a standard generator matrix such as a
Reed--Solomon code or a fountain code \cite{raptor_code}. Let $P_f[t]$
denote the information packet that arrives at the source of flow $f$ in
slot $t$. A  packet $P_f[t]$ of flow $f$ has $k_f$ symbols, the encoded
packet $C_f[t]$ of which is of size $n_f = k_f/r_f$ with $0 < r_f\le1$, and we
assume that the code is such that the packet $P_f[t]$ can be
reconstructed from \emph{any} of its $k_f$ encoded symbols (this is
possible, for example, by Reed--Solomon codes).

\begin{figure}[t]
\begin{center}
{\input{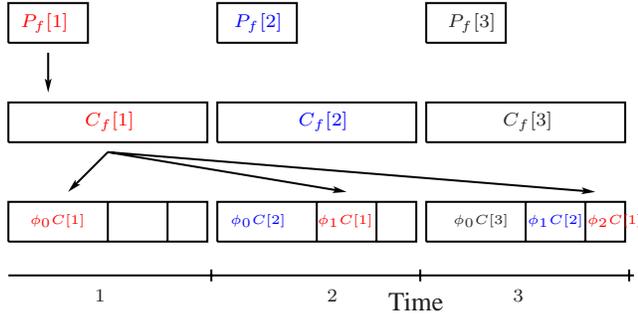}}
\end{center}
\caption{Two stage encoding (example of $D_f = 3)$: information packet
$P_f[1]$ of size $k_f$ is 
encoded to $C_f[1]$ of size $n_f = k_f/r_f$, the contents of which are
allocated across subpackets $\phi_0C_f[1],\phi_1C_f[1],\phi_2C_f[1]$
across $3$ timeslots.}
\label{Fig:coding}
\end{figure}

The second stage allocates the content of the encoded packet $C_t$ of
the first stage across the \emph{transmitted} packets. Each encoded
packet is segmented into $D_f$ portions (where we recall that $D_f$ is
the decoding deadline requirement for each packet of flow $f$), the size
of the $\Delta$th portion being $\phi_f(\Delta) n_f$, where $\Delta
\in \{0,1,\cdots,D_f-1\}$ and $0 \leqslant \phi_f(\Delta) \leqslant 1$.
At transmission slot $t$, a transmitted packet is assembled from the
$\phi_f(0)$ portion of $C_f[t]$, the $\phi_f(1)$ portion of $C_f[t-1]$, and
so on until the $\phi_f(D_f-1)$th portion of packet $C_f[t-D_f+1]$. This
procedure is illustrated in Figure~\ref{Fig:coding} for $n_f=3$. Note
that the transmitted packet is of size $n_f$ symbols. To decode a
packet $P_f[t]$ of flow $f$, we use the transmitted packets that are
received during the transmission slots $t, t+1, \cdots, t+D_f-1$. Note
that the conventional strategy of transmitting an encoded packet every
transmission slot corresponds to the special case: $\phi_f(0) = 1$ and
$\phi_f(1) = \phi_f(2) = \cdots = \phi_f({D_f-1}) = 0$. We call the
transmission scheme outlined above with general $\phi_{\cdot}(\Delta)$s a {\em
generalised block transmission scheme.}

\subsection{Network Constraints on Coding Rate}
Let $w_{f,c}$ be the PHY rate of transmission of flow $f$ in cell $c$.
For each transmitted packet of flow $f$, each cell $c \in \C_f$ along
its route must allocate at least $\frac{n_f}{w_{f,c}}$ units of time to
transmit the packet (or encoded block). Let $\F_c := \{f \in \F: c \in
\C_f\}$ be the set of flows that are routed through cell $c$. We recall
that the transmissions in any cell $c$ are scheduled in a TDMA fashion,
and hence, the total time required for transmitting packets for all
flows in cell $c$ is given by $\sum_{f \in \F_c} \frac{n_f}{w_{f,c}}$.
Since, for cell $c$, the transmission schedule interval is $T_c$ units
of time, the coding rates $r_f$ must satisfy the schedulability
constraint $\sum_{f \in \F_c} \frac{n_f}{w_{f,c}} \leqslant T_c$.

\subsection{Error Probability -- Upper bound}

\begin{lemma}
\label{lem:pe_bound}
The end--to--end probability of a packet erasure for flow $f$ is
bounded by
\begin{align}
&\hspace{7mm}  \widetilde{e}_f \nn
&=    \ \myprob{\sum_{\Delta=0}^{D_f-1} \phi_f(\Delta)\frac{k_f}{r_f} E_f[\Delta] > n_f-k_f} \nn
  &\leq \ \exp\left(-\left[
  \theta_f(1-r_f)-
\sum_{\Delta=0}^{D_f-1}\ln\left(1-\beta_f +\beta_f e^{ \theta_f\cdot \phi_f(\Delta)} \right)\right]\right)\nn
  &=:   \ e_f.\nonumber
\end{align}
where $\theta_f >0$ is the Chernoff--bound parameter. 
\end{lemma}
Let the random variable $\alpha_f[t]$ indicate whether packet $P_f[t]$
is successfully decoded or not, i.e., 
\begin{eqnarray*}
\alpha_f[t] & = & \left\{ 
\begin{array}{ll}
1, & \text{if packet $P_f[t]$ is decoded successfully} \\
0, & \text{otherwise}.
\end{array}
\right.
\end{eqnarray*}
We note here that the decoding errors for the successive packets are
correlated, as each encoded packet overlaps with the transmission of
previous $D_f-1$ packets and the successive $D_f-1$ packets. Hence, the
sequence of random variables $\alpha_f[1], \alpha_f[2], \alpha_f[3],
\cdots $ are correlated.  But, the probability of any $\alpha_f[t] = 0$
is upper bounded by Lemma~\ref{lem:pe_bound}.

\section{Network Utility Maximisation} 
\label{sec:num}
For flow $f$, the total expected throughput as a result of transmitting
$T \geq 1$ packets is given by

\footnotesize
\begin{eqnarray*}
&& S_f(T) \nn
&=&  \sum_{(x_1,x_2,\cdots,x_T) \in \{0,1\}^T}
\left(\sum_{t=1}^T k_f x_t\right) {\sf P}\left\{\alpha_f[t] = x_t, t=1,2,\cdots,T \right\}
\end{eqnarray*}

\normalsize
Note that the joint probability mass function ${\sf P}\left\{\alpha_f[t]
= x_t, t=1,2,\cdots,T \right\}$ is not a product--form distribution as
the packet erasures $\alpha_f[t]$s are correlated. However, the above
expectation can be written as 
\begin{align*}
S_f(T) 
&= \sum_{t=1}^T \ 
   \sum_{x_t \in \{0,1\}} 
   k_f x_t {\sf P}\left\{\alpha_f[t] = x_t \right\} \\
&= T \cdot k_f \cdot (1-e_f) 
\end{align*}
Thus, the (average expected) flow throughput is defined as 
\begin{align*}
S_f &= \lim_{T\to\infty} \frac{S_f(T)}{T} \\
    &= k_f \cdot (1-e_f). 
\end{align*}
We are interested in
maximising the utility of the network which is defined as the sum
utility of flow throughputs. We consider the log of throughput as the
candidate for the utility function being motivated by the desirable
properties like proportional fairness that it possesses.

We define the following notations: the Chernoff--bound parameters ${\bm
\theta}:=[\theta_f]_{f \in \F}$, coding rates ${\bm r} := [ r_f]_{f \in
\F}$, and the allocation of coded bits across transmission slots ${\bm 
\Phi} := [{\bm \phi}_f]_{f \in \F}$ where ${\boldsymbol \phi}_f := 
[\phi_f(0), \phi_f(1), \cdots, \phi_f(D_f-1)]$. Thus, we define the 
network utility as  
\begin{eqnarray}
\widetilde{U}\left({\bm \Phi}, {\bm \theta}, {\bm r}\right)
& := & \sum_{f \in \F} \ln\left(k_f \left(1 - e_f({\bm \phi}_f,
\theta_f, r_f)\right)\right)\nn
&  =: & \sum_{f \in \F} \ln\left(k_f\right) + U\left({\bm \Phi}, {\bm \theta}, {\bm r}\right)
\end{eqnarray}
The problem is to obtain the optimum coded bit allocation ${\bm
\Phi}^*$, the optimum Chernoff--bound parameter ${\bm \theta}^*$, and
the optimum coding rate ${\bm r}^*$ that maximises the network utility. 
Since, $k_f$, the size of information packets of each flow $f$ is given,
maximising the network utility is equivalent to maximising
$U({\bm \Phi}, {\bm \theta},{\bm r}) := \sum_{f\in \F}
\ln\left(1-e_f\right)$. Thus, we define the following problem

\noindent
{\bf P1:}
\newline
\begin{minipage}{\columnwidth}
\small
\begin{align}
\max_{{\bm \Phi},{\bm \theta},{\bm r}} \ \ \ &
U({\bm \Phi}, {\bm \theta},{\bm r}) \nn
\text{subject to} \quad
&\underset{f: c \in \C_f  }{\sum} \frac{k_f}{r_f w_{f,c}}  \leq  T_c,  &&\forall c \in \C \label{eq:cons1} \\ 
&   \underset{\Delta=0}{\overset{D_f-1}\sum} \phi_f(\Delta) =  1,  &&\forall f \in \F  \label{eq:cons2} \\
& \phi_f(\Delta)  \ge  0,   &&\forall f \in \F, 0\leq\Delta \leq D_f-1  \nonumber\\
 &\theta_f  >  0, &&\forall f \in \F                       \nn
 & r_f  \leq  \overline{\lambda}_f \,  &&\forall f \in \F  \nn
 & r_f  \geq  \underline{\lambda}_f \, &&\forall f \in \F  \nonumber
\end{align}
\normalsize
\end{minipage}

\noindent
We note that the Eqn.~\eqref{eq:cons1} enforces the network capacity (or
the network schedulability) constraint. The objective function $U({\bm
\Phi}, {\bm \theta}, {\bf r})$ is separable in $({\bm \phi}_f, \theta_f,
r_f)$ for each flow $f$. Importantly, the component of utility function
for each flow $f$ given by $\ln\left(1-e_f({\bm 
\phi}_f,\theta_f,r_f)\right)$ is not jointly concave in $({\bm \phi}_f,
\theta_f,r_f)$. However, $\ln\left(1-e_f({\bm
\phi}_f,\theta_f,r_f)\right)$ is concave in each of $\phi_f(\cdot)$,
$\theta_f$, and $r_f$. Hence, the network utility maximisation problem
${\bf P1}$ is not in the standard convex optimisation framework.
Instead, we pose the following problem,

\noindent
{\bf P2:}
\newline
\begin{minipage}{\columnwidth}
\small
\begin{align}
\max_{{\bm \Phi}} \max_{\bm \theta} \max_{\bm r} 
& 
\sum_{f \in \F} \ln\left(1-e_f({\bm \phi}_f,\theta_f,r_f)\right) \label{eqn:coordinate_max_problem} \\
\text{subject to} \quad
&\underset{f: c \in \C_f  }{\sum} \frac{k_f}{r_f w_{f,c}}  \leq  T_c,  &&\forall c \in \C \nonumber \\ 
&   \underset{\Delta=0}{\overset{D_f-1}\sum} \phi_f(\Delta) =  1,  &&\forall f \in \F   \nonumber\\
& \phi_f(\Delta)  \ge  0,   &&\forall f \in \F, 0\leq\Delta \leq D_f-1  \nonumber\\
 &\theta_f  >  0, &&\forall f \in \F                       \nn
 & r_f  \leq  \overline{\lambda}_f \,  &&\forall f \in \F  \nn
 & r_f  \geq  \underline{\lambda}_f \, &&\forall f \in \F  \nonumber
\end{align}
\normalsize
\end{minipage}

\noindent
In general, the solution to ${\bf P2}$ need not be the solution to ${\bf
P1}$. However, in our problem, we show that ${\bf P2}$ achieves the solution
of ${\bf P1}$.
\begin{lemma}
\label{lem:opt_joint_vs_sep}.
For a function $f:{\cal Y}\times{\cal Z}\to {\mathbb R}$ that is concave
in $y$ and in $z$, but not jointly in $(y,z)$, the solution to the joint
optimisation problem for convex sets ${\cal Y}$ and ${\cal Z}$ 
\newline
\begin{minipage}{\columnwidth}
\begin{align}
\label{eqn:joint_opt_prob}
\max_{y \in {\cal Y}, z \in {\cal Z}} f(y,z) 
\end{align}
\end{minipage}
is the same as
\newline
\begin{minipage}{\columnwidth}
\begin{align}
\max_{z \in {\cal Z}} \max_{y \in {\cal Y}} f(y,z), 
\end{align}
\end{minipage}
\newline
if $f(y^*(z),z)$ is a concave function of $z$, where for each $z \in
{\cal Z}$, $y^*(z) := \underset{y \in {\cal
	Y}}{\arg\max} f(y,z)$.
\end{lemma}

We note that for each $r_f$ and $\theta_f$, the probability of error
$e_f$ is convex in ${{\bm \phi}_f}$, and hence, $\ln(1-e_f)$ is concave
in ${\bm \phi}_f$. Thus, we first solve for the optimum code bit
allocation ${\bm \phi}^*_f$ in Section~\ref{subsec:opt_Phi}. Then, using
the optimum code bit allocation, we solve for the optimum Chernoff bound
parameter ${\bm \theta}^*$ which we describe in subsection
\ref{subsec:opt_theta}. After having solved for the optimum ${\bm
\theta}^*$, we show in Section~\ref{subsec:convexity} that $U({\bm
\Phi}^*, {\bm \theta}^*({\bm r}), {\bm r})$ is a concave function of
${\bm r}$. Hence, from Lemma~\ref{lem:opt_joint_vs_sep}, the solution to
problem $({\bf P2})$ (the maximisation problem that separately obtains
the optimum ${\bm \theta}^*$ and optimum ${\bm r}^*$) is globally
optimum. We study the rate optimisation problem that obtains ${\bm r}^*$
in subsection \ref{subsec:opt_R}. 

\section{Utility Optimum Rate Allocation} 
\label{sec:sol}

\subsection{Optimal Code Bit Allocation $\boldmath{\Phi}$} 
\label{subsec:opt_Phi}
We consider the maximisation problem defined in 
Eqn.~\ref{eqn:coordinate_max_problem} for a given coding rate vector ${\bm r}$
and Chernoff--bound parameter vector ${\bm \theta}$, and  
obtain the optimum ${\bm \phi}_f$ for each flow $f \in \F$.
The sub--problem is given by
\begin{eqnarray*}
\max_{{\bm \phi}_f} & & \sum_{f \in \F} \ln\left(1-e_f({\bm \phi}_f,\theta_f,r_f)\right)\\
\text{subject to} & &  
\begin{array}{lllll}
\underset{\Delta=0}{\overset{D_f-1}\sum} \phi_f(\Delta) & = & 1, &
\forall f \in \F & \\
 \phi_f(\Delta) & \ge & 0,  & \forall f \in \F, \forall \Delta
\leq D_f-1. &
\end{array}
\end{eqnarray*}
This is a separable convex optimisation problem, and hence can be solved
by Lagrangian method. Let $\mu_f$ be a Lagrangian multiplier for the
constraint $\underset{\Delta=0}{\overset{D_f-1}\sum} \phi_f(\Delta)  =
1$, and define ${\boldsymbol \mu} = [\mu_f]_{f \in \F}$. The Lagrangian
function is given by
\begin{eqnarray*}
L({\bf \Phi}, \boldsymbol \mu)
 = \sum_{f \in \F} \ln\left(1-e_f\right)
- \sum_{f \in \F} \mu_f \bigg(1 - \sum_{\Delta=0}^{D_f-1}
  \phi_f(\Delta)\bigg) 
\end{eqnarray*}
Applying KKT condition, 
\begin{eqnarray*}
\frac{\partial L \hspace{4mm} }{\partial \phi_f(i)}\mid_{\phi_f(i)^*} &=& 0,
\end{eqnarray*}
we get 
\begin{eqnarray}
\label{eqn:kkt_opt_phi}
0 &= &\frac{-e_f}{1-e_f} \cdot
\frac{\beta_f \theta_f e^{\theta_f {\phi^*_f(i)}}}{1-\beta_f + \beta_f
e^{\theta_f \phi^*_f(i)}} + \mu_f \nonumber\\
\text{or}, \ 
e^{\theta_f {\phi^*_f(i)}} &=&\frac{1-\beta_f}{\beta_f} 
\frac{(1-e_f)\mu_f}{\theta_fe_f-\mu_f(1-e_f)}
\end{eqnarray}
for $i = 0,1,2,\cdots,n_f-1$. Since, the RHS of	
Eqn.~\ref{eqn:kkt_opt_phi} is the same for all $i$, we  
get $\phi^*_f(i) = \phi^*_f(j)$, and hence 
\[
\phi^*_f(\Delta) = \frac{1}{D_f}, \ \forall \Delta = 0,1,\cdots,D_f-1.
\]
Thus, ${\bf \Phi}^*$ allocates equal portions of an encoded packet
across transmission schedules with a delay of $0,1,\cdots, D_f-1$,
unlike the conventional transmission scheme which transmits all the
coded bits of a packet in one shot. Hence, 
$e_f({\bm \phi}_f^*,\theta_f,r_f)$ is 
\begin{align}
\label{eqn:e1_bound}
e_f &= \exp\left(-\left[
\theta_f (1-r_f) - D_f \ln\left(1-\beta_f+\beta_f e^{\frac{\theta_f}{D_f}}\right)
\right]\right).
\end{align}

\subsection{Optimal $\boldmath{\theta}^*$}
\label{subsec:opt_theta}
We now consider the optimum Chernoff--bound parameter problem with the
optimum coded bits allocation ${\bm \Phi}^*$, and for any given coding
rate vector ${\bm r} \in
[\underline{\lambda}_f,\overline{\lambda}_f]^F$. 

\begin{minipage}{\columnwidth}
\begin{align}
\max_{\boldsymbol \theta} 
& 
\sum_{f \in \F} \ln\left(1-e_f({\bm \phi}_f^*,\theta_f,r_f)\right) \label{eqn:max_theta_problem} \\
\text{subject to} \quad
 &\theta_f  >  0, \ \ \ \ \  \forall f \in \F \nonumber                       
\end{align}
\end{minipage}

\vspace{4mm}
We note that the objective function is separable in $\theta_f$s, and
that $e_f$ is convex in $\theta_f$. Hence, the problem defined in 
Eqn.~\eqref{eqn:max_theta_problem}, is a concave maximisation problem.
The partial derivative of $e_f$ with respect to $\theta_f$ is given by
\begin{align*}
\frac{\partial e_f}{\partial\theta_f}
&=  -e_f \cdot \left[ (1-r_f) - \frac{\beta_f
e^{\theta_f/D_f}}{1-\beta_f+\beta_fe^{\theta_f/D_f}}\right]. 
\end{align*}
Observe that $\frac{\beta_f
e^{\theta_f/D_f}}{1-\beta_f+\beta_fe^{\theta_f/D_f}}$ is an
increasing function of $\theta_f$.
Thus, if, for $\theta_f = 0$, $1-r_f - \frac{\beta_f }{1-\beta_f+\beta_f}
< 0$ or $r_f > 1-\beta_f$, the
derivative is positive for all $\theta_f >0$, or $e_f$ is an increasing
function of $\theta_f$. Hence, for $r_f > 1-\beta_f$, the optimum
$\theta_f^*$ is arbitrarily close to $0$ which yields $e_f$
arbitrarily close to $1$. Thus, for error
recovery, for any end--to--end
error probability $\beta_f$, the coding rate should be smaller than
$1-\beta_f$, in which case, we
obtain the optimum $\theta_f^*$ by equating 
the partial
derivative of $e_f$ with respect to $\theta_f$ to zero.
\begin{align*}
\begin{array}{lrcl}
\text{i.e.,}&  \frac{\beta_f
e^{\theta_f^*/D_f}}{1-\beta_f+\beta_fe^{\theta_f^*/D_f}} & = & 1-r_f \\
\text{or},  & e^{\theta_f^*/D_f} & = &
\frac{1-r_f}{\beta_f}\frac{1-\beta_f}{r_f} \\
\text{or},  & \theta_f^* & = & D_f \left[\ln\left(\frac{1-r_f}{\beta_f}\right) -
 \ln\left(\frac{r_f}{1-\beta_f}\right)\right]. 
\end{array}
\end{align*}
Thus, the probability of a packet decoding error for a given $r_f$ with
the optimum allocation of coded bits ${\bm \Phi}^*$, and the optimum
Chernoff--bound parameter  $\theta_f^*$, is 
\begin{align*}
& \ \ \ \ e_f\\
&=  \exp\left(-D_f\left[
(1-r_f) \ln\left(\frac{1-r_f}{\beta_f}\right) +r_f
\ln\left(\frac{r_f}{1-\beta_f}\right)
\right]\right)\\
&= \exp\left(-D_f \cdot \text{KL}(\mathcal{B}(1-r_f)||\mathcal{B}(\beta_f)\right)
\end{align*}
where KL$(f_1,f_2)$ is the Kullback--Leibler divergence between the
probability mass functions (pmfs) $f_1$ and $f_2$. 

\subsection{A convex optimisation framework to obtain optimal $r_f^*$}
\label{subsec:convexity}
If $\ln(1-e_f({\bm \phi}^*_f,\theta_f^*,r_f))$ is concave in $r_f$, then
one can obtain the optimum $r_f^*$ using convex optimisation framework.
To show the concavity of $\ln(1-e_f({\bm \phi}^*_f,\theta_f^*,r_f))$ it
is sufficient to show that $e_f({\bm \phi}^*_f,\theta_f^*,r_f)$ is
convex in $r_f$. Note that
\begin{align*}
\frac{\partial e_f}{\partial r_f} &= e_f \cdot \theta_f^*(r_f)\\
\frac{\partial^2 e_f}{\partial r_f^2} 
 &= 
{e_f}
 \left[
 \theta_f^{*2} - \frac{D_f}{r_f(1-r_f)} 
 \right]
\end{align*}
$e_f$ is convex if 
\begin{align*}
\left[\ln\left(\frac{1-r_f}{\beta_f}\right) -
\ln\left(\frac{r_f}{1-\beta_f}\right)\right]^2
&\geq 
 \frac{D_f}{r_f(1-r_f)},
\end{align*}
or, 
\begin{align*}
\ln\left(\frac{1-r_f}{r_f} \frac{1-\beta_f}{\beta_f} \right)
&\geq 
 \frac{\sqrt{D_f}}{\sqrt{r_f(1-r_f)}}\\
 \text{or}, \ 
 \frac{\sqrt{D_f}}{\sqrt{r_f(1-r_f)}} - 
\ln\left(\frac{1-r_f}{r_f} \frac{1-\beta_f}{\beta_f} \right)
&\leq 0 
\end{align*}
The function $\frac{1}{\sqrt{r_f(1-r_f)}}$ is convex in $r_f$.
Also, $\ln\left(\frac{1-r_f}{r_f}\right)$ is decreasing with $r_f$, and
hence, 
$-\ln\left(\frac{1-r_f}{r_f}
\frac{1-\beta_f}{\beta_f} 
\right) \leq 
-\ln\left(\frac{1-\overline{\lambda}_f}{\overline{\lambda}_f}
\frac{1-\beta_f}{\beta_f} \right)$. 
Thus, we have a
sufficient condition
\begin{align}
\label{eqn:constraint}
 \frac{\sqrt{D}_f}{\sqrt{r_f(1-r_f)}} - 
\ln\left(\frac{1-\overline{\lambda}_f}{\overline{\lambda}_f} \frac{1-\beta_f}{\beta_f} \right)
&\leq 0 
\end{align}
The above condition requires the delay deadline $D_f$ to be smaller than
some $\overline{D}_f(r_f)$. We consider $D_f$s to satisfy this
condition, and hence, the rate optimisation problem is a concave
maximisation problem. For the sake of completeness, we include this as a
constraint in the problem formulation. However, this condition is not an 
active constraint.

\subsection{Optimal Coding Rate $\boldmath{r}$}
\label{subsec:opt_R}
From the previous subsection, we observe under the delay constraint
Eqn.~\eqref{eqn:constraint} that $e_f({\bm \phi}^*_f,
\theta_f^*(r_f),r_f)$ is convex in $r_f$, and hence, we obtain the
optimum coding rate $r_f^*$ using convex optimisation method.  Also,
from Lemma~\ref{lem:opt_joint_vs_sep}, it is clear that $r_f^*$ is the
unique globally optimum rate. Thus, we solve the following network
utility maximisation problem 
\newline
\begin{minipage}{\columnwidth}
\begin{align}
\max_{\boldsymbol r} 
& 
\sum_{f \in \F} \ln\left(1-
e_f({\bm \phi}^*_f, \theta_f^*(r_f),r_f)
\right) \label{eqn:max_x_problem} \\
\text{subject to} \quad
&\underset{f: c \in \C_f  }{\sum} \frac{k_f}{r_f w_{f,c}}  \leq  T_c,  &&\forall c \in \C \nonumber \\ 
 & r_f  \leq  \overline{\lambda}_f \,  &&\forall f \in \F  \nn
 & r_f  \geq  \underline{\lambda}_f \, &&\forall f \in \F  \nn
 &  \frac{\sqrt{D_f}}{\sqrt{r_f(1-r_f)}} - a
 \leq 0 \, &&\forall f \in \F
 \label{eqn:max_x_problem_k_constraint} 
\end{align}
\end{minipage}
\newline
where $a = 
\ln\left(\frac{1-\overline{\lambda}_f}{\overline{\lambda}_f}
\frac{1-\beta_f}{\beta_f} \right)$.
It is clear that the objective function is separable and concave, and
hence, can be solved using Lagrangian relaxation method. Also, we note
here that the constraint represented
by Eqn.~\eqref{eqn:max_x_problem_k_constraint} is not an active constraint,
and hence, there is no Lagrangian cost to this constraint.
We note here
that the coding rate should be such that $k_f/r_f$ is an integer,
and hence, obtaining $r_f^*$ is a discrete
optimisation problem. This is, in general, an NP hard problem. Hence, we relax
this constraint, and allow $r_f$ to take any real value in  
$[\underline{\lambda}_f, \overline{\lambda}_f]$. 
The
Lagrangian function for the rate optimisation problem is thus
\begin{align*}
& \ \ \ \ L({\boldsymbol r},{\boldsymbol p},
{\boldsymbol u},
{\boldsymbol v} 
)\\
&= 
\sum_{f \in \F} \ln\left(1 - e_f\right)
- \sum_{c \in \C}p_c\left(\sum_{f \in \F_c} \frac{k_f}{r_f w_{f,c}}
  - T_c\right)\\
  &
+ \sum_{f \in \F}u_f\left(r_f - \underline{\lambda}_f\right) 
- \sum_{f \in \F}v_f\left(r_f - \overline{\lambda}_f\right) 
\end{align*}
Applying KKT condition,
$\frac{\partial L}{\partial r_f}\mid_{r_f^*} = 0$,
we have
\begin{align*}
\frac{-1}{1-e_f} \frac{\partial e_f}{\partial r_f}\mid_{r_f^*} 
&= \sum_{c \in \C_f}p_c \frac{-k_f}{r_f^{*2}w_{f,c}} + v_f - u_f\\ 
&= 
\frac{-k_f}{r_f^{*2}} \left( \sum_{c \in \C_f}\frac{p_c}{w_{f,c}}
\right) + v_f - u_f\\ 
%
 \frac{e_f}{1-e_f}\cdot \theta_f^*
&=
\frac{k_f}{r_f^{*2}} \left( \sum_{c \in \C_f}\frac{p_c}{w_{f,c}}
\right) + v_f - u_f.
\end{align*}
If the optimum $r_f^*$ is either $\underline{\lambda}_f$ or 
$\overline{\lambda}_f$, then it is unique. If 
$r_f^* \in (\underline{\lambda}_f, \overline{\lambda}_f)$, then $u_f
= v_f = 0$, 
which is the most
interesting case, and we consider only this case for the rest of the paper. 
Let  $\lambda_f :=   \sum_{c \in \C_f}\frac{p_c}{w_{f,c}}$. The above
equation becomes
\begin{align}
\frac{e_f}{1-e_f} \cdot \theta_f^* 
&= \frac{\lambda_f k_f}{r_f^{*2}} \label{eq:lambda} \\
e_f &=
							  \frac{\lambda_fk_f}{\lambda_fk_f+\theta_f^*r_f^{*2}
							  }\label{eqn:ef}\\ 
\exp\left(-D_f D({\cal B}(1-r^*_f)\|{\cal B}(\beta_f))\right)
&=
                              \frac{\lambda_fk_f}{\lambda_fk_f+\theta_f^{*}r_f^{*2}
							  }\nn 
D_f D({\cal B}(1-r_f^*)\|{\cal B}(\beta_f))
&=
\ln\left(\frac{\lambda_fk_f+\theta_f^*r_f^{*2}}{\lambda_fk_f}\right)
\label{eqn:x_star}
\end{align}
In the above equation, the LHS is a strictly convex decreasing function
of $r_f^*$. Since, the utility maximisation problem is a concave
maximisation problem, the optimum rate $r_f^* \in (0, 1-\beta_f)$ exists
and is unique.

\subsection{Sub--gradient Approach to Compute optimum ${p}_c^*$}
In this section, we discuss the procedure to obtain the Shadow costs or
the Lagrange variables ${\bm p}^*$. The dual problem for the primal
problem defined in Eqn.~\eqref{eqn:max_x_problem} is given by
\begin{eqnarray*}
\min_{{\bm p} \geq 0} & & D(\bm{p}),
\end{eqnarray*}
where the dual function $D(\bm{p})$ is given by
\begin{align}
& \hspace{5mm} D(\bm{p})\nn
&= \max_{{\bm r}} \underset{f\in\F}{\sum}\ln(1-e_f(r_f))+\underset{c\in\C}{\sum}p_c\left( T_c - \underset{f \in \F_c}{\sum}
\frac{k_f}{r_f w_{f,c}}
\right) \label{eq:supremum}\\
&= \underset{f\in\F}{\sum}\ln(1-e_f(r_f^*({\bm
p})))+\underset{c\in\C}{\sum}p_c\left( T_c - \underset{f \in \F_c}{\sum}
\frac{k_f}{r_f^*({\bm p})w_{f,c}} 
\right). \\ \nonumber
 \label{eq:d_star}
\end{align}
\normalsize

\noindent
In the above equation, $e_f(r_f)$ denotes $e_f({\bm \phi}^*_f,
\theta_f^*(r_f),r_f)$. Since the dual function (of a primal problem) is
convex, $D$ is convex in ${\bm p}$. Hence, we use a sub--gradient method
to obtain the optimum $\bm{p}^*$.
From Eqn.~\eqref{eq:supremum}, it is clear that for any ${\bm r}$, 
\begin{align*}
D(\bm{p}) &\geq \underset{f\in\F}{\sum}\ln(1-e_f(r_f))+\underset{c\in\C}{\sum}p_c\left( T_c - \underset{f \in \F_c}{\sum}
\frac{k_f}{r_fw_{f,c}}
\right),
\end{align*}
\normalsize
and in particular, 
$D({\bm p})$ is greater than that for $r=r_f^*({\widetilde{\bm p}})$, i.e.,
\begin{align}
& \hspace{5mm}D(\bm{p})\nn
&\geq \underset{f\in\F}{\sum}\ln(1-e_f(r_f^*({\widetilde{\bm p}})))+\underset{c\in\C}{\sum}p_c\left( T_c - \underset{f \in \F_c}{\sum}
\frac{k_f}{r_f^*({\widetilde{\bm p}})w_{f,c}}
\right)\nn
&= D(\widetilde{\bm p})
+\underset{c\in\C}{\sum}\left(p_c-\widetilde{p}_c\right)\left( T_c - \underset{f \in \F_c}{\sum}
\frac{k_f}{r_f^*({\widetilde{\bm p}})w_{f,c}}
\right)
\end{align}
\normalsize
Thus, a sub--gradient of $D(\cdot)$ at any $\widetilde{\bm p}$ is given
by the vector
\begin{align}
 \left[ T_c - \underset{f \in \F_c}{\sum}
\frac{k_f}{r_f^*({\widetilde{\bm p}})w_{f,c}}\right]_{c \in \C}.
\end{align}
We obtain an iterative algorithm based on sub--gradient method that
yields $\bm{p}^*$, with ${\bm p}(i)$ being the Lagrangians at the $i$th
iteration. 
\begin{eqnarray*}
p_c(i+1) = \left[p_c(i)-\gamma\cdot
\left(T_c - \underset{f \in \F_c}{\sum}
\frac{k_f}{r_f^*({\bm p}(i))w_{f,c}} 
\right)\right]^+.
\end{eqnarray*}
where $\gamma > 0$ is a sufficiently small stepsize, and $[f(x)]^+ := \max\{f(x),0\}$
ensures that the Lagrange multiplier
never goes negative. Note that the Lagrangian updates can be locally
done, as each cell $c$ is required to know only the rates $r_f^*({\bm
p}(i))$ of flows $f \in \F_c$. Thus, at the beginning of each iteration
$i$, the flows choose their coding rates to $r_f^*({\bm p}(i))$, and each cell
computes its cost based on the rates of flows through it. The updated
costs along the route of each flow are then fed back to the source node
to compute the rate for the next iteration.

The Lagrange multiplier $p_c$ can be viewed as the cost of transmitting
traffic through cell $c$. The amount of service time that is available
is given by $\delta = T_c - \underset{f \in \F_c}{\sum}
\frac{k_f}{r_f^*({\bm p}(i))w_{f,c}}$. When $\delta$ is positive and
large, then the Lagrangian cost $p_c$ decreases rapidly (because $D$ is
convex), and when $\delta$ is negative, then the Lagrangian cost $p_c$
increases rapidly to make $\delta \geq 0$. We note that the increase or
decrease of $p_c$ between successive iterations is proportional to
$\delta$, the amount of service time available. Thus, the sub--gradient
procedure provides a dynamic control scheme to balance the network loads.

We explore the properties of the optimum rate parameter $r_f^*$ in 
Section~\ref{subsec:x_f_s}. In Section~\ref{sec:examples}, we provide
some examples that illustrate the optimum utility--fair resource
allocation.

\subsection{Properties of $r_f^*$}
\label{subsec:x_f_s}

\begin{lemma}
\label{lem:monotone_x_star}
$r_f^*(D_f)$ is an increasing function of $D_f$.
\end{lemma}

Lemma~\ref{lem:monotone_x_star} is quite intuitive. For any given
channel error $\beta_f$, as the deadline become less stringent, it
is optimal to go for a high rate code. In other words, it is optimal for
a flow to use as much scheduling time as possible (for a large
$D_f$, and hence, use a high rate code); however, the
resources are shared among multiple flows, and hence, we ask the
following question:  ``{\em what is the optimal share of the scheduling
time}'' that each flow should have. Interestingly, in our problem
formulation, the code rate $r_f$ also solves this optimal scheduling
times for each flows.

\section{Examples}
\label{sec:examples}

\subsection{Example 1: Two cells with equal traffic load}

\begin{figure}[t]
\begin{center}
\includegraphics[width=50mm, height=35mm]{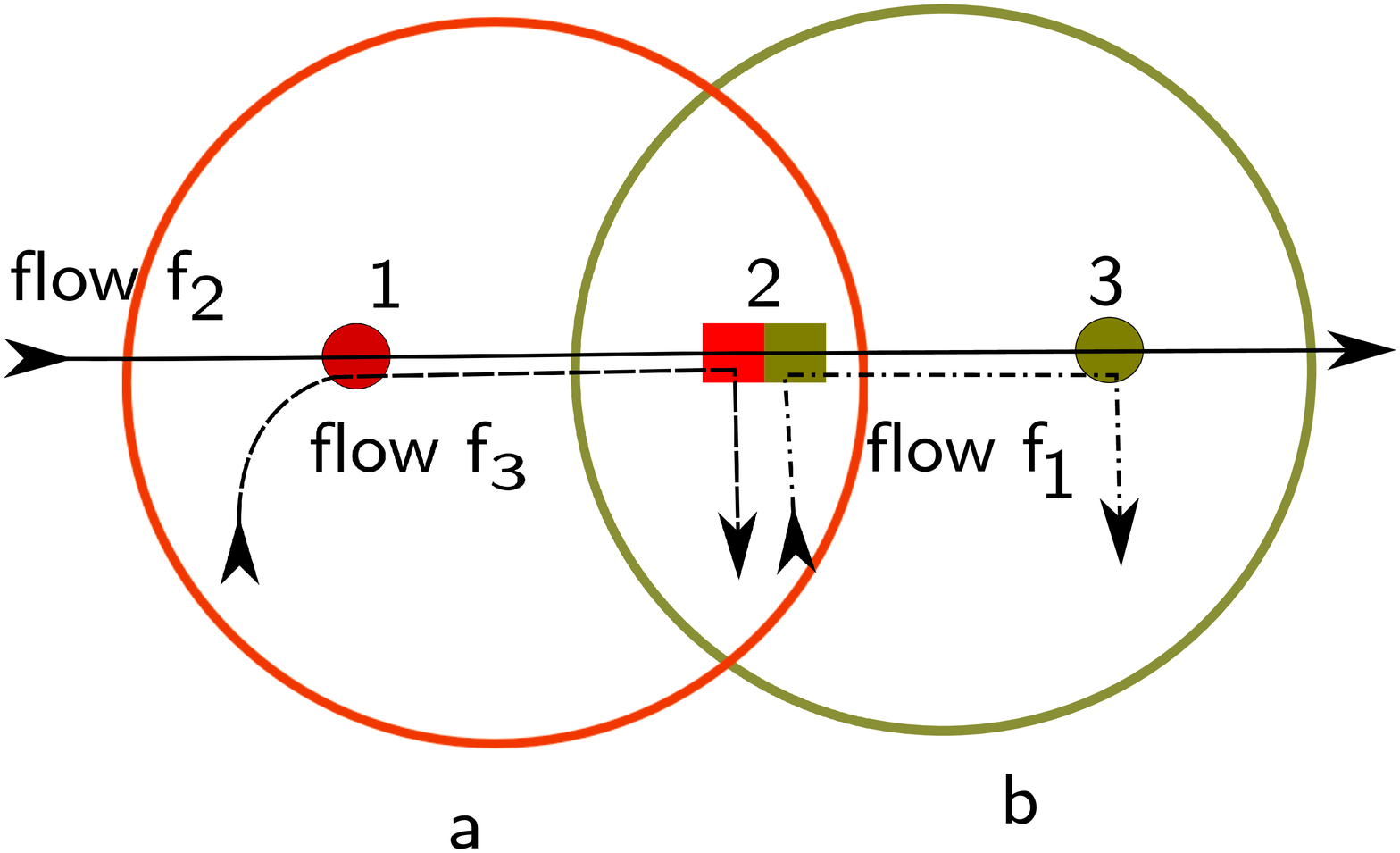}
\caption{Cells with equal traffic load}
\label{fig:example_equal_traffic}
\end{center}
\end{figure}

We begin by considering the example shown in 
Figure~\ref{fig:example_equal_traffic} consisting of two cells $a$ and
$b$ having three nodes 1, 2, and 3. Each cell has the same packet
erasure probability $\beta$ and the schedule length $T$.  There are
three flows $f_1, f_2$, and $f_3$, with two of the flows $f_1$ and
$f_3$ having one--hop routes $\C_{f_1} = \{b\}$ and $\C_{f_3} = \{a\}$,
and one flow $f_2$ having a two--hop route $\C_{f_2} = \{a,b\}$. Each
flow has the same information packet size $k$, decoding deadline $D$ and
PHY transmit rate, \emph{i.e.} $w_{f,c} = w$. This is analogous to the
so--called parking--lot topology often used to explore fairness issues. 

The end--to--end erasure probability experienced by the two--hop flow
$f_2$ is greater than that experienced by the one hop flows $f_1$ and
$f_3$, since each hop has the same fixed erasure probability. Hence, we
need to assign a lesser coding rate $r_{f_2}$ to flow $f_2$ than to
flows $f_1$ and $f_3$ in order to obtain the same error probability
(after decoding) across flows. However, when operating at the boundary
of the network capacity region (thereby maximising throughput),
decreasing the coding rate $r_{f_2}$ of the two--hop flow $f_2$ requires
that the coding rate of \emph{both} one--hop flows  $f_1$ and $f_3$ be
increased in order to remain within the available network capacity. In
this sense, allocating coding rate to the two--hop flow $f_2$ imposes a
greater marginal cost on the network (in terms of the sum--utility) than
the one--hop flows, and we expect that a fair allocation will therefore
assign higher coding rate to the two--hop flow $f_2$. The solution
optimising this trade--off in a proportional fair manner can be
understood using the analysis in the previous section.     

In this example, both the cells are equally loaded and, by symmetry, the
Lagrange multipliers $p_a = p_b$. Hence, $\lambda_{f_1} =
\frac{\lambda_{f_2}}{2} = \lambda_{f_3}$. For the Chernoff--bound 
parameter ${\boldsymbol \theta} = [\theta, \theta]$, we find
from Eqn.~\eqref{eq:lambda}, 
\begin{eqnarray*}
\frac{e_{f_2}}{1-e_{f_2}} \cdot
\frac{1-e_{f_1}}{e_{f_1}}
&=& 
\frac{\lambda_{f_2}}{\lambda_{f_1}}
\cdot \frac{r_{f_1}^{*2}}{r_{f_2}^{*2}} \nn
& = & 2 \cdot \frac{r_{f_1}^{*2}}{r_{f_2}^{*2}}.
\end{eqnarray*}
For sufficiently small erasure probabilities, we have
\begin{eqnarray*}
\frac{e_{f_2}}{e_{f_1}}
& \approx & 2 \cdot \frac{r_{f_1}^{*2}}{r_{f_2}^{*2}}\nn
& \approx & 2 
\end{eqnarray*}
Thus the
proportional fair allocation is $e_{f_1} = e_{f_3}
\approx 1/2\cdot e_{f_2}$. That is, the
coding rates are allocated such that the one--hop flows have
approximately half the error probability of the two--hop flow.    

\subsection{Example 2: Two cells with unequal traffic load}

We consider the same network as in the previous example, but now with only the 
flows $f_1$ and $f_2$ (i.e., the flow $f_3$ is not present) in the
network. In this example, cell b carries two flows while cell a carries only one flow. The
encoding rate constraints are given by
\begin{eqnarray*}
\frac{1}{r_{f_2}}  & \leq & \frac{wT}{k}, \ \text{(from cell 1)}, \\ 
\frac{1}{r_{f_1}} + \frac{1}{r_{f_2}} & \leq & \frac{wT}{k}, \ \text{(from cell 2)}.
\end{eqnarray*}
Since, both $r_{f_1}$ and $r_{f_2}$ are at most 1, it is clear that at
the optimal point, the rate constraint of cell a is not tight while the
constraint of cell b is tight. Thus, the shadow prices (Lagrange
multipliers) $p_{a} = 0$ and $p_{b} >0$. That is, at the
first hop the cell is not operating at capacity, and so the ``price''
for using this cell is zero. In this example, $\lambda_{f_1} =
\lambda_{f_2}$, and hence, from Eqn.~\eqref{eq:lambda}, we deduce that for
sufficiently low cell erasure probability $\beta$, $e_{f_1}
\approx e_{f_2}$. Alternatively, as the delay deadline $D \to \infty$, from 
Eqn.~\eqref{eq:lambda} we have $e_{f_1} = e_{f_2}$.
These proportional fair allocations make sense intuitively since
although flow $f_2$ crosses two hops, it is only constrained at the
second hop and so it is natural to share the available capacity of this
second hop approximately equally between the flows. When the erasure
probability is sufficiently small, this yields approximately the same
error probabilities for both flows. For larger erasure probabilities,
it leads to the two--hop flow having higher error probability, in
proportion to the per--hop erasure probability $\beta$.

\begin{figure}[t]
\centering
\includegraphics[width=50mm, height=35mm]{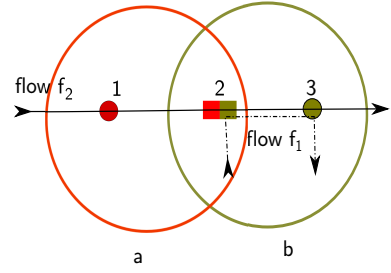}
\caption{Cells with unequal traffic load}
\label{fig:example_unequal_traffic}
\end{figure}
\bibliographystyle{IEEEtran}
\bibliography{IEEEabrv,allerton}
\end{document}